\documentclass[useAMS,usenatbib]{mn2e}
\usepackage[dvips]{graphicx}

\title[The suppressed dynamical friction in a core]{The test for
suppressed dynamical friction in a constant density core of dwarf
galaxies}
\author[S. Inoue]{Shigeki
Inoue$^{1}$\thanks{E-mail:inoue@astr.tohoku.ac.jp}
\\
$^{1}$Astronomical Institute, Tohoku University, Sendai 980-8578, Japan}
\begin{document}

\date{2009 January 29}

\pagerange{\pageref{firstpage}--\pageref{lastpage}} \pubyear{2009}

\maketitle

\label{firstpage}

\begin{abstract}
The dynamical friction problem is a long-standing dilemma about globular
 clusters (hereafter,GCs) belonging to dwarf galaxies. GCs are
 strongly affected by dynamical friction in dwarf galaxies, and are presumed
 to fall into the galactic center. But, GCs do exist in dwarf galaxies
 generally. 
A solution of the problem has been proposed. If dwarf
 galaxies have a core dark matter halo which has constant density
 distribution in its center, the effect of dynamical
 friction will be weakened considerably, and GCs should be able to survive
 beyond the age of the universe. Then, the solution argued that, in a
 cored dark halo, interaction between the halo and the GC constructs a
 new equilibrium state, in which a part of the halo rotates along with
 the GC (co-rotating state). The equilibrium state can suppress the dynamical friction in the core region.
In this study, I tested whether the solution is reasonable and
 reconsidered why a constant density, core halo suppresses dynamical
 friction, by means of N-body simulations.
As a result, I conclude that the true mechanism of suppressed dynamical
 friction is not the co-rotating state, although a core halo can
 actually suppress dynamical friction on GCs significantly.
\end{abstract}


\section{Introduction}

The world we live in is a hierarchical universe, in which galaxies are made by a myriad of merging events. A large-scale numerical simulation based on the Cold Dark Matter (CDM) theory has been operated as a greatly declarative method for the hierarchical scenario. It indicated that the structure of the universe develops from the small dark matter clumps which collapsed first, and result in the formation of large and massive dark mater halos. In such a formation history, it is appropriate to consider that dwarf galaxies are fundamental 'building-blocks' and expected to be the oldest structures of the universe. Dwarfs are believed to have important clues in understanding the hierarchical universe.

In this paper, I will discuss the dynamical friction problem which refers to orbital motion of GCs in dwarf galaxies. The drag force of dynamical friction is negligibly weak for GCs in the Milky Way. In contrast, it operates strongly in small systems like dwarf galaxies (see chap. 8 of \citealt{b7}). Thus, the GCs in dwarfs are presumed to lose their orbital energy and fall into the galactic center by strong friction force from the dark matter halo. According to results of both analytical and numerical studies, the timescale for a GC to fall into the center is of the order of $\sim 1 Gyr$ \citep{Tremaine,HernandezGilmore,OhLinRicher,Vesperini00,Vesperini01,Sanchez,Goerdtetal}. Nevertheless, even in the present universe, these GCs still do exist and keep their orbital motions. For example, the \textit{Fornax} dSph galaxy has five GCs which are metal poor and as old as the universe, thus resembling the GCs of Milky Way \citep{Buonannoetal98,Buonannoetal,Straideretal,MackeyGilmore,Grecoetal}.

However, by an analytical approach, \citet{HernandezGilmore} have discovered that a King model halo can significantly weaken the effect of dynamical friction in the core region. As for a cuspy halo (NFW profile or singular isothermal sphere), a GC is sucked into the galactic center by the dynamical friction (see the fig.2 of \citet{Goerdtetal}). The analytical approach of \citet{HernandezGilmore} was constructed on the Chandrasekhar dynamical friction formula \citep{Chandra}. On the other hand, by N-body simulations, \citet{Goerdtetal} and \citet{Readetal} (hereafter, R06) confirmed the cessation of dynamical friction on a GC in a core region of halos. However, R06 concluded that an important key to this suppressed dynamical friction is a `co-rotating state'. They argued that a part of halo particles in the constant density core begin to rotate with the GC. The authors suggested that this dynamical state is a new equilibrium state including the GC. The dynamical friction ceases under this equilibrium; hence, the GC could survive beyond the age of the universe. \citet{Goerdtetal} and R06 confirmed that these results don't depend on the mass of a GC, the orbital parameters (circular or elliptical orbit) of a GC, or the size of core structure of a halo (core radius).

However, the conclusions of these studies, \citet{HernandezGilmore} and R06, imply a discrepancy between them. The approach of \citet{HernandezGilmore} is based on the Chandrasekhar formula; hence, it cannot take account of velocity anisotropy of the field particles, because the formula is based on the assumption of isotropic velocity state. But, contrary, R06 concluded that the mechanism of suppressed dynamical friction is the very anisotropy in the velocity state: the co-rotating state. My aim in this paper is to assess the authenticity of the co-rotating state proposed by R06 as the mechanism of the suppressed dynamical friction.

Although the result of R06 appears to be convincing, the new equilibrium, the co-rotating state, will be vulnerable to perturbation on the GC. Once the GC orbit is perturbed and the orbital plane inclined, the co-rotating equilibrium state will be broken easily. This means that the system is no longer equilibrium: the dynamical friction force on the GC will be rejuvenated. R06 have also mentioned this fragile nature of the co-rotating equilibrium state. In R06 they studied single GC cases only. But, real dwarf galaxies don't necessarily have only one GC, but several or more \citep{Durrelletal,Milleretal,Lotzetal}. The GCs in a dwarf will be perturbed by the other GCs. Then, I infer that some GCs would fall into the galactic center by such orbital perturbation, and may merge and form a stellar nucleus cluster at the galactic center \citep{Miocchietal,CapuzzoMiocchi}. Actually, some dwarf galaxies have a nucleus stellar cluster at the center, and some observational researchers have discussed that some of these nuclei may be remnants of GCs \citep{Milleretal,Lotzetal}. I conducted N-body simulations to examine whether the GCs in dwarfs fall into the galactic center by perturbation from the other GCs, even in the case of cored halo structure.
 
This paper is organized as follows: in the next section, I will explain my simulation method and models in detail; in the 3rd section I will show my simulation results and their analysis; I will discuss my results, comparing them with preceding studies, and give my conclusion in the 4th section.

\section{The simulations}
My simulational settings are almost the same as the N-body simulation of R06. The simulations are pure N-body simulations (no gas component). I use a Barnes-Hut modified tree-code \citep{BarnesHut, Barnes} in order to lighten the heavy burden of gravitational force calculation, setting an open angle of $\theta = 0.5$. A special-purpose calculator for collisionless N-body simulations, GRAPE-7 model 600, is used with the tree algorithm to accelerate gravity calculation \citep{Makino}. The total number of timesteps is 11841 for the whole of a simulated period which corresponds to 10 Gyr in real timescale. It takes roughly a half month to finish each simulation. 
 
\subsection{The setting of halo model}
To imitate R06, I adopt the same spherical density distribution:
\begin{equation}
 \rho(r)=\frac{\rho_0}{(r/r_s)^\gamma\bigl[1+(r/r_s)^\alpha\bigr]^{(\beta-\gamma)/\alpha}}
  \quad\qquad ;\;r<r_{vir}
\label{density}
\end{equation}
with $\alpha = 1.5$, $\beta = 3.0$, $\gamma = 0.0$. The density in the core $\rho_0$ is $0.10 M_{\odot}pc^{-3}$. The scale radius $r_s$ is set to $0.91 kpc$. The density is nearly constant at the center within 200-300 pc, which defines the core region. The virial mass of the halo is $2.0\times10^9M_{\odot}$. I add an exponentially decaying envelope to prevent instability at the outer region caused by an artificial cut-off radius $r_{vir}$ \citep{SpringelWhite}. Velocity dispersion of particles is given by the solution of Jeans equations as a function of radius,
\begin{equation}
 \sigma_r^2(r)=\frac{1}{r^{2\beta}\rho(r)}\int_r^{\infty}dr'r'^{2\beta}\rho(r')\frac{d\Phi}{dr'},
 \label{jeans}
\end{equation}
where $\beta$ is the anisotropy parameter. Although the effect of dynamical friction is sensitive to the velocity distribution of the field particles, isotopic velocity state is supposed to be reasonable in inner region of dwarf halos \citep{Mashchenko}. In this paper, I assume the isotropic velocity state in the halo, setting $\beta=0$ ($\sigma_r=\sigma_\theta=\sigma_\phi$). With this assumption, Eq.\ref{jeans} reduces to
\begin{equation}
 \sigma^2(r)=\frac{1}{\rho(r)}\int_r^{\infty}dr'\rho(r')\frac{d\Phi}{dr'}.
 \label{iso}
\end{equation}
The velocity distribution is determined by the local Maxwellian approximation,
\begin{equation}
 F(v,r)=4\pi\Bigl(\frac{1}{2\pi\sigma^2}\Bigr)^{3/2}v^2\exp{\Bigl(-\frac{v^2}{2\sigma^2}\Bigr)},
 \label{maxwellian}
\end{equation}
where $F(v,r)$ is a probability distribution function of velocity \citep{Hernquist}. Eq.\ref{maxwellian} is normalized so that $\int_0^\infty F(v,r)dv=1$.

Like R06 and \citet{Goerdtetal}, I adopt a three-shell model \citep{Zempetal}, which consists of finer grained particles in inner regions and coarser particles in outer regions. This technique enables it to reduce computational run-time, and resolve much smaller scales in the inner region. But, this multi-shell model inevitably admits the heavier particles coming from the outer shells into the inner shells, and may induce two-body relaxation between these different mass particles. To avoid such an unfavorable artificial effect, I refine the particles in the outer shells depending on their orbit. From an set of initial position and velocity, I can calculate the pericenter distance of a specific particle in the smooth potential given by the density profile, Eq.\ref{density}. By the pericenter distance, I detect the heavier particles which are supposed to intrude into the inner region. I divide these intruding particles into a set of particles which have the same mass resolution as the particles in the inner region. The new particles will have the same radial velocity component as the original particle but a new random tangential component of the same magnitude as the original one. The divided particles are randomly placed on a sphere whose radius is the same as the initial galactocentric distance of the original particle. For a simple explanation, let's suppose that the halo consists of shells A, B and C, from inner to outer shell. The shells A, B and C consists of particles which have the mass of $m_A$, $m_B$ and $m_C$, respectively ($m_A<m_B<m_C$). For example, if a certain particle which has the mass of $m_C$ and an initial position in the shell C intrudes into the shell A, the particle will be divided into a set of new particles which have the mass of $m_A$, and the number of the new particles will be $m_C/m_A$ (for detail, see \citealt{Zempetal}); therefore, not all particles in a outer shell have a uniform mass resolution, but two-body relaxation can be minimized in the inner most region. In the simulations of R06 or \citet{Goerdtetal}, they have been missing the particle-dividing step; therefore, at this point my simulations have an improvement over the preceding studies.

Specifically, the particle masses are $m_A=17.8M_{\odot}$, $m_B=356M_{\odot}$ and $m_C=7118M_{\odot}$. The inner most region, the shell A, is within $300 pc$ (to be accurate, the shell A is not a shell, but a sphere). The shell B, the middle shell, is the region from $300-1100 pc$. The shell C, the outer most shell, is the entire region of outside of the shell B. The number of particles is $8.31\times10^6$ in total. The softening lengths of the particles, $m_A$, $m_B$ and $m_C$, are 3, 8 and 22 pc, respectively. I checked that my results were not sensitive to these values.

The three-shell model has coarse resolution in the outer shells. Such heavyer particles may affect the nature of dynamical friction. For confirmation, I ran a simulation with an uniform mass particle model, and checked that the results were not sensitive to these model settings (see Appendix A). 
 
By analytical calculation, \citet{Sanchez} has proposed that the dynamical friction induced by the stellar component is not negligible in a cored dwarf halo. But, the purpose of this paper is to judge the co-rotating state proposed by R06 as the mechanism of the suppressed dynamical friction. For fair comparison with the simulation of R06 in which stellar component was excluded, I don't take the effect from stellar component into account in my simulations here.

\subsection{The setting of globular clusters}
In my simulations, each GC is represented by a point mass with $m_{GC}=2.0\times10^5M_{\odot}$. The softening length is $10 pc$. Just to make sure, I ran another simulation in which a GC was resolved by many particles, and confirmed that tidal disruption didn't destroy the GC. In this study, I don't consider mass-loss from a GC, merging between GCs, or dynamical heating by halo potential. Although these effects may play important roles in the case of the resolved GCs \citep{FujiiFunatoMakino,Miocchietal,EsquivelFuchs,CapuzzoMiocchi}, I consider the GCs as point-masses for the sake of comparison with R06 or \citet{Goerdtetal}.

\section{the results}
In this paper, I operate simulations for the cases of 1, 5 and 30 GCs. In single GC cases, I give the GC a set of specific orbital parameters in each case. On the other hand, in multi GC cases, I set their orbits at basically random as detailed in the following subsections.

\subsection{The single GC cases}

\begin{figure}
 \includegraphics[width=84mm]{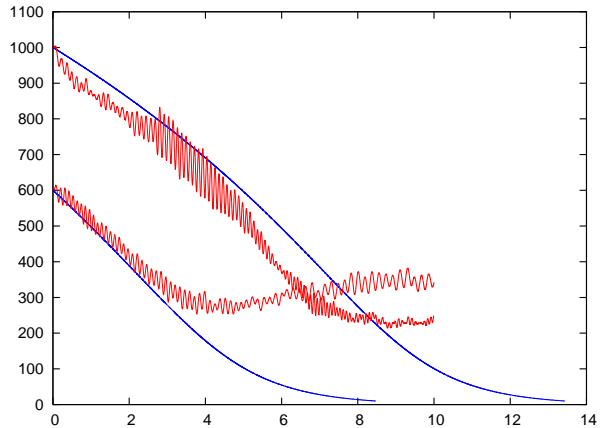}
 \caption{The comparison between analytical and numerical results of time evolution of orbital radius of a GC. The smooth lines indicate the analytical results, the waving jaggy lines indicate the N-body results. I estimated the analytical dynamical friction with $\ln\Lambda=3.0$. The initial orbit of the GC is a circular orbit from radius 600 pc or 1 kpc.}
 \label{Chan}
\end{figure}
\begin{figure}
 \includegraphics[width=84mm]{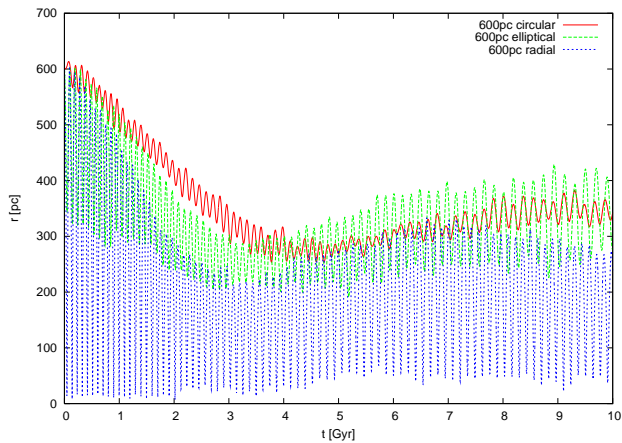}
 \caption{The results of single GC cases, each line indicates different simulation. The red line represents the case in which a GC is placed on the circular orbit of radius 600 pc initially. The green, blue lines are for the GC initially placed on elliptical and radial orbit from radius 600 pc, respectively. The core region is inside 200-300 pc, in all simulations here.}
 \label{fig1}
\end{figure}

To begin with, I conduct some single GC cases. Initial orbit of the GC is circular and set to 600 pc or 1 kpc. In Fig.\ref{Chan}, I show two comparison cases with the Chandrasekhar dynamical friction formula (the derivation of the analytical result is described in Appendix B .). As seen in the figure, the analytical results correspond to the N-body results before the GC enters into the core region (inside 200-300 pc). But, when entering into the core, these results diverge abruptly: the analytic results continue to fall into the galactic center, while the orbital shrinkage by dynamical friction stops in the N-body cases. 

However, as I noted above, the effect of dynamical friction is sensitive to the velocity distribution of the field particles. But I confirmed that the core stalling of dynamical friction was not sensitive to details of the velocity distribution function with various anisotropy parameters $\beta$ in Eq.\ref{jeans}.  

\begin{figure}
 \includegraphics[width=84mm]{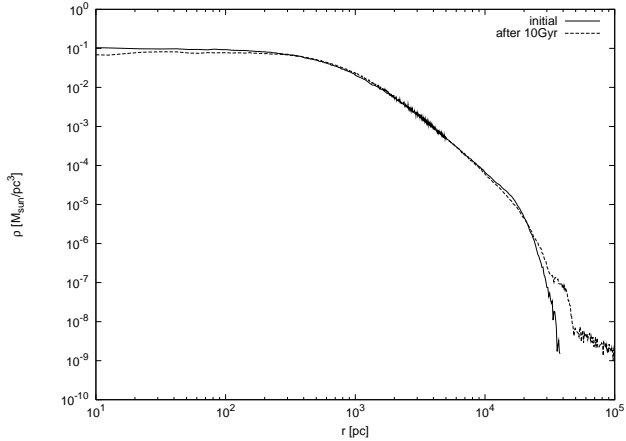}
 \caption{The density profiles in the case of the circular orbit with 1 kpc initial orbital radius (the upper jaggy line in Fig.\ref{Chan}). The solid line is the initial state, and the dashed line indicates the end of the simulation (after 10 Gyr). The horizontal axis is distance from the galactic center, the ordinate is mass density of the halo, $\rho [M_{\odot}/pc^3]$.}
 \label{fig2}
\end{figure}

Next, I investigate the relation of the suppressed dynamical friction with orbital eccentricities. The initial orbit is circular, elliptical or radial and set to 600 pc. In the case of elliptical orbit, the rotational velocity of the GC is initially set to $0.6v_c$ ($v_c$ is the circular velocity at the initial position) and the radial velocity is $0$. In the case of radial orbit, the GC is at rest initially. The results are indicated in Fig.\ref{fig1}. As shown in the figure, when the GC enters into the core region, the orbital shrinkage by dynamical friction stops in all cases regardless of their initial orbital eccentricities. After the cessation of dynamical friction, the orbit expands a little again. This phenomenon is called `kickback effect', and a detailed investigation about the effect has been done by \citet{Goerdtetal2008}. In this study, I don't treat this effect. These behaviors of the GC in the cored profile (the cessation of orbital shrinkage, the kickback effect) are consistent with R06 and \citet{Goerdtetal}. The mass included in the halo core is heavier than a GC by two orders, which should be enough to operate dynamical friction. \citet{Goerdtetal} has confirmed independence of suppressed dynamical friction from the size of the core region, $r_s$. Moreover, I find that the density profile of the halo is scarcely changed by the GC (Fig.\ref{fig2}). The energy conservation rate of the system, $1-E_{end}/E_{ini}$, is $6.46\times10^{-3}$. 

\begin{figure}
 \includegraphics[width=84mm]{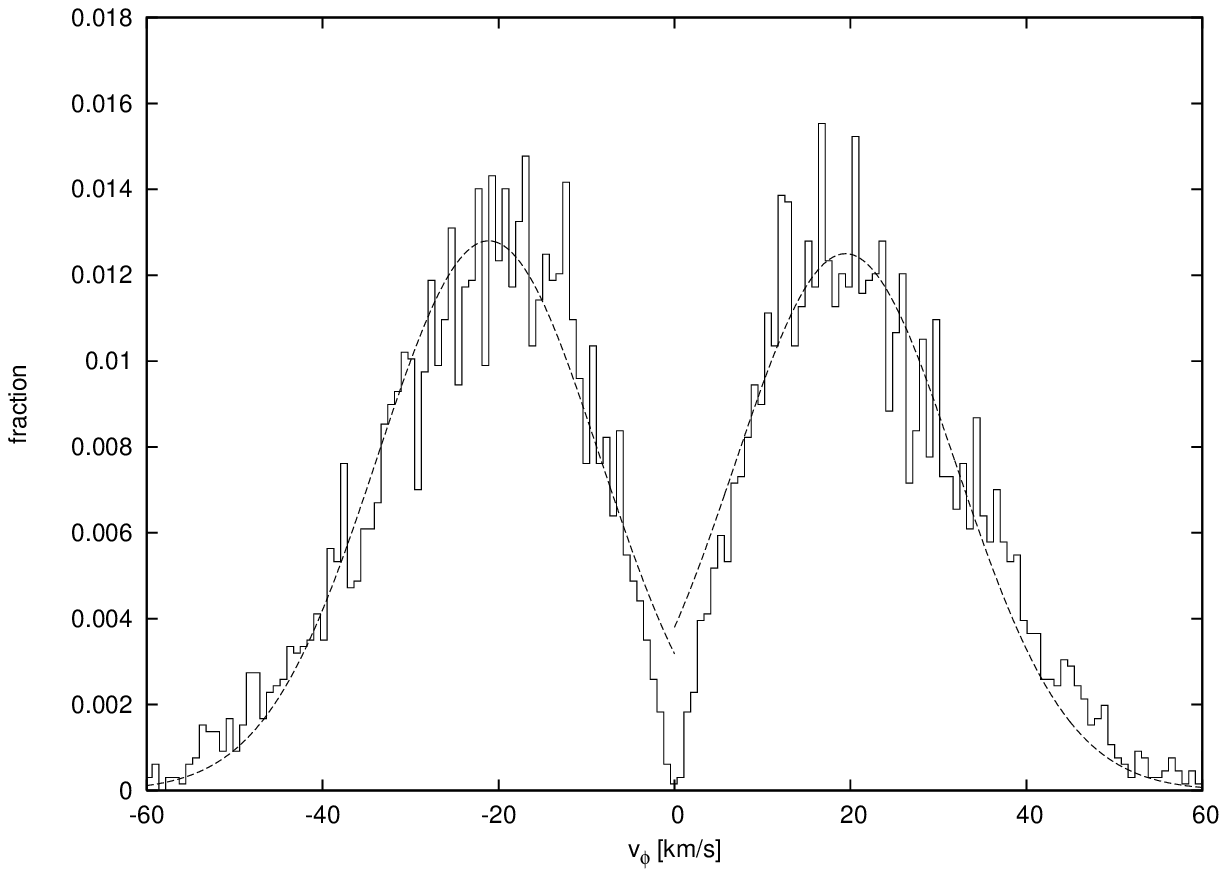}
 \includegraphics[width=84mm]{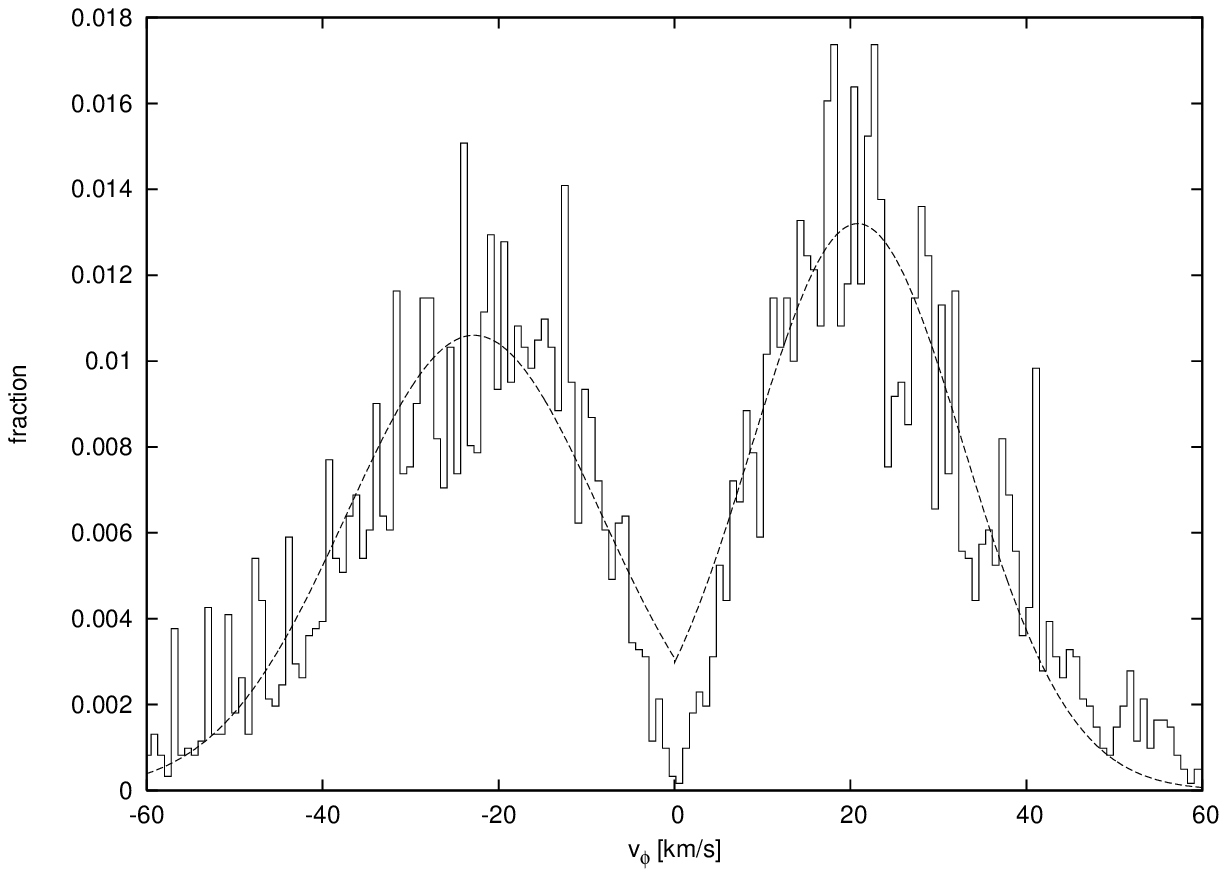}
 \caption{The histograms of the distribution of azimuthal velocity $v_\phi$ for the halo particles satisfying the condition of Eq.\ref{eq2} and $150pc<r_p<300pc$. The vertical axis is mass fraction. The upper panel is for the initial state, and the bottom panel is for $t = 8.2 Gyr$. The positive side of $v_\phi$ represents pro-grade rotating particles with the GC, the negative side represents retro-grade motion. The dashed lines are Gaussian fitting for each side.}
 \label{fig3}
\end{figure}

This result confirms the suppressed dynamical friction on a GC in a cored halo. However, what is the cause of it? In R06, the authors have proposed that a part of halo particles in core region are made to rotate with the GC by gravitational interaction (see the fig.4 in R06). They called this dynamical state `co-rotating state' and concluded it to be the mechanism to suppress the dynamical friction on the GC. I will examine whether the co-rotating state is the true mechanism or not. To identify the co-rotating state, I follow the same manner as R06. In order to visualize the velocity state of the halo, I sieve out the particles which have the radial distance $r_p=150-300 pc$ , subject to a condition, 
\begin{equation}
\frac{|\mathbf{J}_p\cdot\mathbf{J}_c|}{|\mathbf{J}_p||\mathbf{J}_c|}>\cos
 \theta,
 \label{eq2}
\end{equation}
where $\mathbf{J}_{p,c}$ means the specific angular momentum of a halo particle and the GC, respectively. A criterion parameter $\theta$ is set to $10^\circ$, which is the same value as R06. Because the halo potential is spherically symmetric and the GC is significantly heavier than any halo particles, the direction of $\mathbf{J}_c$ is assumed to be constant in time. The condition, Eq.\ref{eq2}, screens out the field particles for which the direction of angular momentum vector coincides with that of the GC within $\theta$. Fig.\ref{fig3} indicates histograms of rotational velocity distributions in the case of a circular orbit for which the initial orbital radius is 1 kpc (the upper jaggy line in Fig.\ref{Chan}). The upper panel of Fig.\ref{fig3} shows the initial state, the bottom panel is for $t = 8.2 Gyr$ (after the cessation of dynamical friction). As seen in the bottom panel, the co-rotating state is constructed after the dynamical friction is suppressed. The velocity distribution becomes somewhat anisotropic: the fraction of pro-grade rotating particles seems to increase, whereas retrograde particles decrease. The figure is consistent with the result of R06. The over-plotted dashed lines in the histograms represent Gaussian fitting given by the minimum $\chi^2$ method. In the bottom panel (co-rotating state), the peak height of the fitting line for pro-grade side $h_+$ and the retrograde side $h_-$ is 0.0132 and 0.0106, respectively. The residual fraction of these, $(h_+-h_-)/h_-$, is 0.245. This value means that the peak of the pro-grade side is 24.5 per cent higher than that of the retrograde side. The minimum value of $|\chi|$ for the bottom panel is 0.00175, and $|\chi|/h_-$ is 0.165.

\begin{figure}
 \includegraphics[width=84mm]{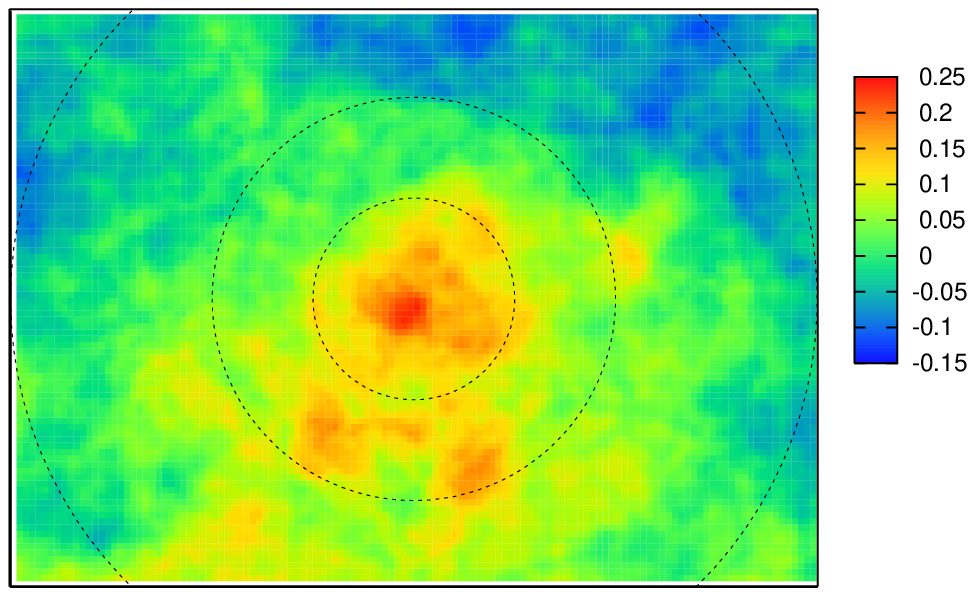}
 \includegraphics[width=84mm]{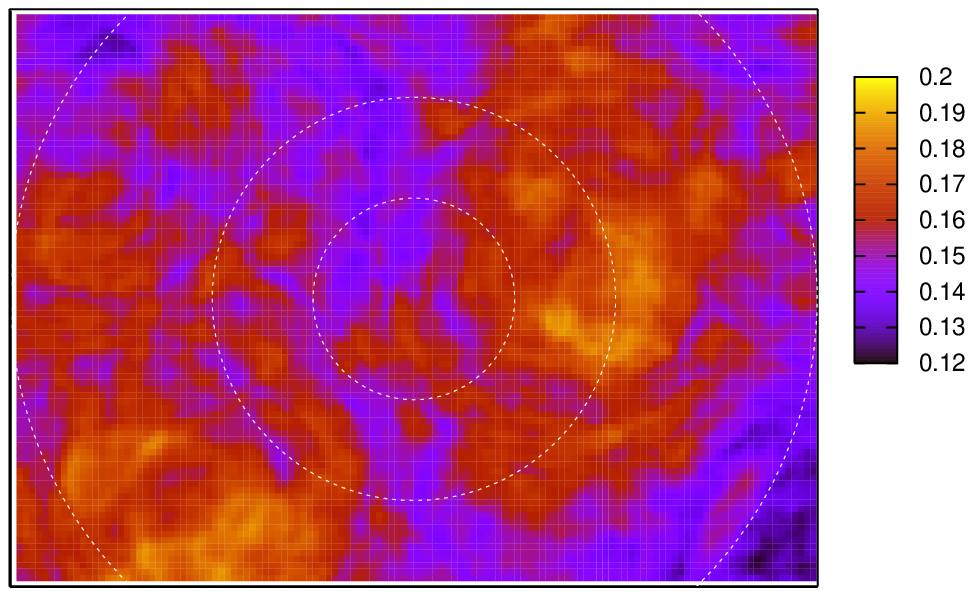}
 \caption{The contour maps of $(h_+-h_-)/h_-$ (upper) and $|\chi|/h_-$ (bottom). The center is the direction for which the value of $(h_+-h_-)/h_-$ is the largest (`the maximal direction'). The dotted circles indicate 5$^\circ$, 10$^\circ$ and 20$^\circ$ from this direction, from inside to outside.}
 \label{fig4}
\end{figure}

From this analysis, the existence of co-rotating state seems to be confirmed. But it may be marginal because the value of $(h_+-h_-)/h_-$ is not significantly larger than $|\chi|/h_-$. One point to note is that the direction of $\mathbf{J}_c$, which is assumed to be constant, actually fluctuates due to the N-body nature of this simulation (\textrm{i.e.}, a finite number of particles has been used) For the sake of more precise discussion, I evaluate $(h_+-h_-)/h_-$ and $|\chi|/h_-$ for slightly different directions of $\mathbf{J}_c$. I re-analyze the velocity states, changing the inclination of the vector $\mathbf{J}_c$ in Eq.\ref{eq2} little by little. By this procedure, I draw contour maps of the value of $(h_+-h_-)/h_-$ and $|\chi|/h_-$. Fig.\ref{fig4} indicates the results. As shown in the figure, the co-rotating state ($(h_+-h_-)/h_->0$) can be found within $\sim 10^\circ$ from the maximal direction.
But, the value of $(h_+-h_-)/h_-$ should be compared with the value of $|\chi|/h_-$. The contour map of $|\chi|/h_-$ is given in the bottom panel of Fig.\ref{fig4}. From the map, it is found that the range of the value is $0.12<|\chi|/h_-<0.2$ for the entire area plotted. Therefore, the value of $(h_+-h_-)/h_-$ is almost submerged under the fitting error, $|\chi|/h_-$, in most of the area except the central region. This discussion means that the co-rotating state shown in the upper panel of Fig.\ref{fig4} is statistically marginal and unimportant, maybe except the central region ($\sim3^\circ$ from the maximal direction). Such weak anisotropy couldn't be expected to affect the dynamical friction on a GC.

\begin{figure}
 \includegraphics[width=84mm]{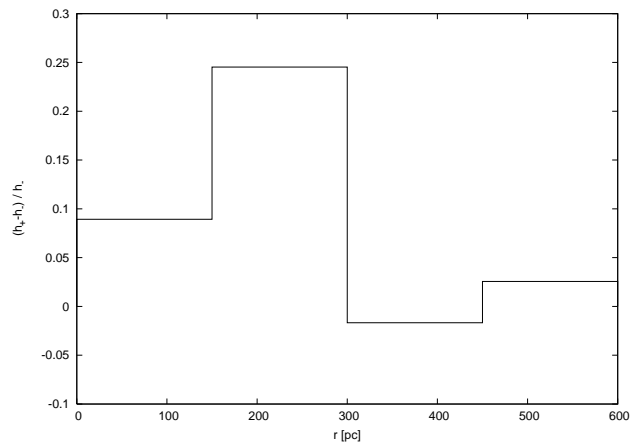}
 \caption{The radial dependence of $(h_+-h_-)/h_-$. The width of each bin represents a radial range of the analysis. $\mathbf{J}_c$ is fixed to the maximal direction.}
 \label{fig5}
\end{figure}

Finally, I inspect the co-rotating state for the dependence on radial distance from the galactic center. So far, I've analyzed the co-rotating state in the radial range of $150pc<r<300pc$. I additionally carry out the same analysis in other radial ranges, $r<150pc$, $300pc<r<450pc$ and $450pc<r<600pc$, with $\mathbf{J}_c$ unchanged. Fig.\ref{fig5} indicates the results. From the figure, it can be seen that the co-rotating velocity state is constructed only in $150pc<r<300pc$.

\subsection{The 5 GCs case}

\begin{figure*}
  \begin{minipage}{150mm}
   \includegraphics[width=160mm]{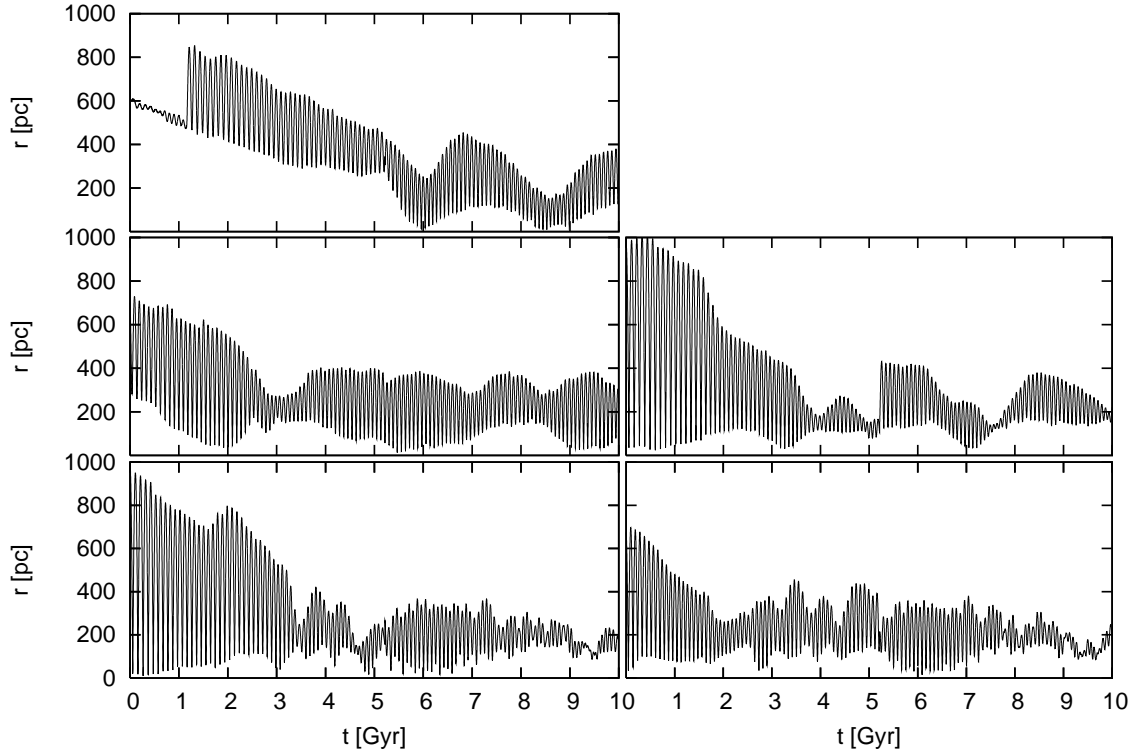}
   \caption{The simulation result of the 5 GCs case, showing the variations of orbital radius. The top left panel represents the reference GC which is initially placed on a circular orbit with a radius 600 pc. The others indicate GCs whose initial distance and velocity are randomly given.}
   \label{fig6}
  \end{minipage}
\end{figure*}

From the discussion of the previous subsection, the co-rotating state not be influential in dynamical friction. Even if the co-rotating state suppresses the dynamical friction on a GC, because it would be fragile against orbital perturbation on the GC as proposed in R06, the dynamical friction would be rejuvenated by the presence of the perturbation. Actually, because real dwarf galaxies generally have some GCs \citep{Milleretal}, there are probably frequent perturbations on the GCs in real dwarfs. As an additional test for the authenticity of the co-rotating state, I perform the simulations of multi GC cases. 

In this subsection, I present the result of the 5GCs case. The GCs are represented as point masses. The initial positions of the GCs are randomly determined in the radial range of $300pc<r<2kpc$. The energy of each GC is also given randomly in the range of $E<\Phi_{2kpc}$ with a random direction of the velocity vector ($\Phi_{2kpc}$ is the potential energy of the halo at $r=2.0 kpc$). To facilitate direct comparison with single GC cases, one GC (reference GC) is placed on a circular orbit with radius 600 pc.

The results for all GCs are shown in Fig.\ref{fig6}. The top left panel indicates the reference GC, and the other panels represent the other 4 GCs which have randomly chosen positions and velocities. As shown in Fig.\ref{fig6}, all orbits of the GCs are frequently perturbed. From the comparison with single GC cases (Fig.\ref{fig1}), the perturbation is expected to be caused by mutual interaction among the GCs, because the difference between these simulations is the number of GCs only. Despite these perturbations, no GCs fall into the galactic center, with all GCs surviving and keeping their orbital motions. This means that the dynamical friction is suppressed in a core region even under frequent perturbation. This implies the inconsistency with the description in R06: the vulnerability of the co-rotatig state to perturbations. This result casts a doubt on the co-rotating state as the mechanism of suppressed dynamical friction, together with the analysis of the single GC case. Actually, the co-rotating state can not be found in velocity histograms of the 5GCs case.

\subsection{The 30 GCs case}

\begin{figure*}
  \begin{minipage}{170mm}
   \includegraphics[width=125mm]{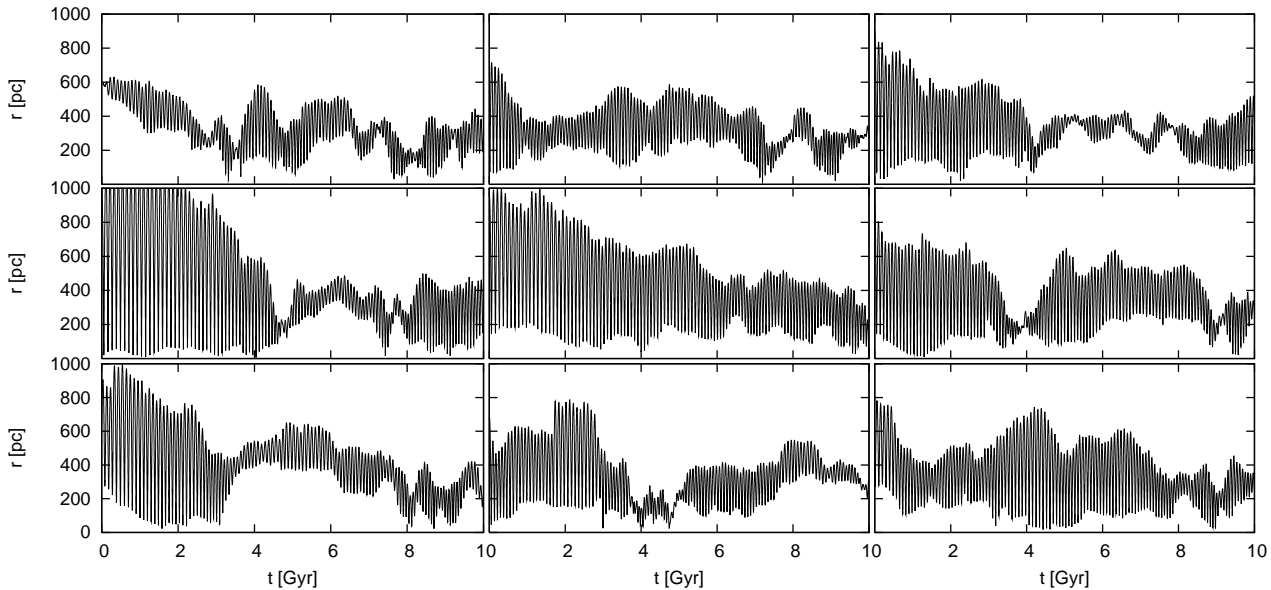}
   \caption{The same as Fig.\ref{fig6}, but for the 30GCs case. The top left panel represents the reference GC.}
   \label{fig7}
  \end{minipage}
\end{figure*}

To confirm the result of the 5GC case, I investigate the case of 30GCs. The number of GCs in this case is somewhat too large, because the actual number of GCs in any dwarf galaxies is $\sim20$ at most \citep{Milleretal}. The settings for the initial condition is basically the same as the 5GCs case: random positions and velocities are given except for one GC which is placed on a circular orbit with radius 600 pc (reference GC). Some simulation results of these GCs are shown in Fig.\ref{fig7}.

As seen from Fig.\ref{fig7}, these results are essentially the same as 5GCs case. Dynamical friction is suppressed in the core region despite more frequent perturbation than the 5GC case. Moreover, the co-rotation state is not confirmed in this case also, and can be guessed to be broken by perturbation among numerous GCs. The result reinforces my argument.

\section{Discussion \& Conclusion}
My simulation results indicate as follows. On one hand, dynamical friction is indeed suppressed in a constant density core region, orbital shrinkage of a GC stops and the orbital motion is sustained. But, on the other hand, this result doesn't depend on the number of GCs. This means that the dynamical friction doesn't work in a core, even if the co-rotating state is broken by frequent perturbation. Moreover, in single GC cases the co-rotating state seems to be too marginal to affect the dynamical friction. My main conclusion here is that the co-rotating state is not the true mechanism. There may be another reason why dynamical friction ceases in a constant density halo.

However, I confirmed that a cored halo structure certainly can weaken the dynamical friction force on a GC in the core region. This means that the cored structure can be the solution of the dynamical friction problem. Actually, as an observational fact, the rotation curves of Low Surface Brightness (LSB) galaxies seem to be like a solid body rotation which means a nearly constant density distribution, although CDM cosmological N-body simulations indicate cuspy density distributions \citep{b5,b6}. Moreover, by the cosmological SPH simulation, \citet{Mashchenko} has discovered a dwarf galaxy which has a constant density core in the halo. They concluded that massive stars inject large amount of energy into the dark matter halo via supernova explosions, and the feedback induced by bursty star formation can turn the cuspy density distribution of the dark matter into a cored profile in the halo center. The suppressed dynamical friction would reinforce the existence of a core structure in dwarfs. If it's the case that the actual dwarfs have a core region in the center, all GCs belonging to a dwarf are included within the core region of the dwarf. Thus, it can be expected that in a dwarf the largest galactocentric distance of the GCs indicates the minimum value of the core radius. For instance, in the \textit{Fornax} dSph, the furthest GC from the galactic center has a projected distance of $\sim 1.6kpc$ \citep{MackeyGilmore}. The core region of the \textit{Fornax} would be larger than $\sim 1.6kpc$.

The most fundamental examination of dynamical friction is embodied by the Chandrasekhar dynamical friction formula \citep{Chandra}. However, several approximations have been dared in the derivation. Current astrophysicists realized that the formula is not always correct because of complex nonlinear effects\citep{JiangBinney,HashimotoFunatoMakino,FujiiFunatoMakino,KimKimSanchez}. Probing the dark matter distribution of dwarfs requires more sophisticated approaches to the nature of dynamical friction.

\section*{Acknowledgments}
The numerical simulations reported here were carried out on GRAPE systems kindly made available by CfCA (Center for Computational Astrophysics) at the National Astronomical Observatory of Japan. The numerical code was based on a software distributed on the website of Joshua E. Barnes (http://ifa.hawaii.edu/\~{}barnes/software.html). I thank Masafumi Noguchi for his fascinating and helpful discussion.

\appendix

\section[]{The reasonability of the three-shell model}
The three-shell model I used here is a complex model. To make sure that the heavier particles with large force sofening lengths in the outer shells don't affect the dynamical friction unfavorably, I run a simulation with an uniform mass resolution model. The uniform mass model consists of particle $m_B$ described in Sec.2.1. The total number of particles is $6.57\times10^6$. The velocity state of the model is isotropic (Eq\ref{iso}).  

\begin{figure}
 \includegraphics[width=84mm]{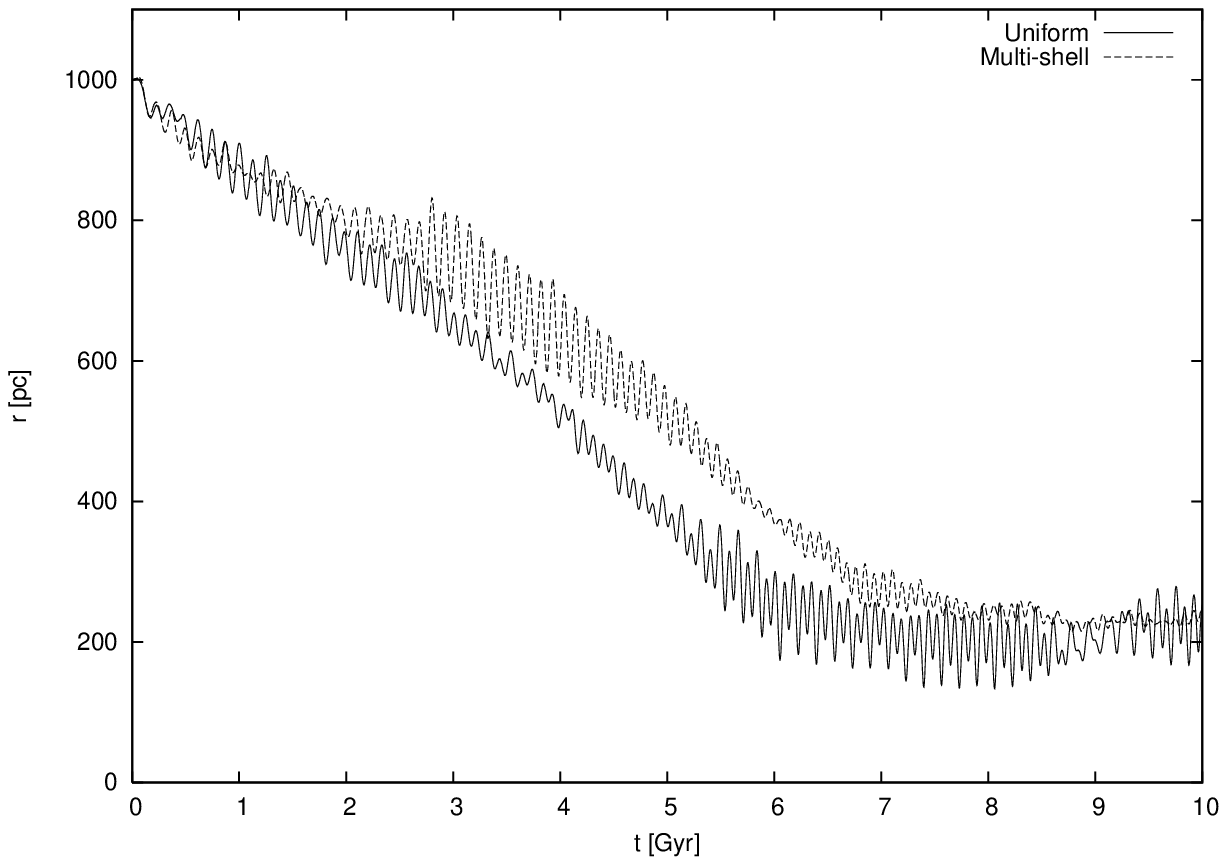}
 \caption{The comparison of the uniform mass model with the three-shell model. The result of the three shell model is the same as the case of the circular orbit from 1 kpc in Sec.3.1.}
 \label{apfig}
\end{figure}

The result is shown in Fig.\ref{apfig}. As seen, there is no difference between the two models in time evolution of the GC orbit and the stalling of dynamical friction in the core region.

\section[]{The derivation of the analytical dynamical friction}
The Chandrasekhar dynamical friction formula is 
\begin{equation}
\frac{dv_{c}(r)}{dt}=-\frac{4\pi\ln\Lambda G^2\rho(r)M_{GC}}{v_{c}^2}\Bigl[\rm{erf}(X)-\frac{2X}{\sqrt{\pi}}e^{-X^2}\Bigr],
 \label{sekhar1}
\end{equation}
where $X\equiv v_{c}/(\sqrt{2}\sigma(r))$. $M_{GC}$ is the GC mass. $v_{c}(r)$ is circular velocity of the GC as a function of radius. $G$ is the gravitational constant. $\Lambda \equiv \frac{b_{max}V_0}{G(m+M_{GC})}$, where $b_{max}$ is the largest impact parameter, $m$ is a mass of field particles, $V_0$ is a typical velocity of the system. In this paper, I approximate $\ln\Lambda = 3.0$ (see \citet{b7}).

I approximate that the orbit is circular. Therefore, $L=M_{GC}rv_c$ and
\begin{equation}
  \frac{dL}{dt}=M_{GC}v_c\frac{dr}{dt}.
  \label{sekhar2}
\end{equation}
Moreover, because dynamical friction drag force is parallel to $v_c$, and therefore a torque
\begin{equation}
  \frac{dL}{dt}=M_{GC}r\frac{v_c}{dt}.
  \label{sekhar3}
\end{equation}
From these equations, 
\begin{equation}
  \frac{dr}{dt}=\frac{r}{v_c}\frac{v_c}{dt}.
  \label{sekhar4}
\end{equation}
Substituting Eq.\ref{sekhar1} for Eq.\ref{sekhar4}, I calculate the time evolution of a GC orbit by numerical integration of Eq.\ref{sekhar4}.

\end{document}